\documentclass[conference]{IEEEtran}
\IEEEoverridecommandlockouts

\usepackage{subfigure}
\usepackage{booktabs}
\usepackage{amsmath}
\usepackage{bm}
\usepackage{colortbl}
\usepackage{tabularx}
\usepackage{color}
\usepackage{epsfig}
\usepackage{xspace}
\usepackage{graphicx}    
\usepackage{url}         %
\usepackage{amssymb}
\usepackage{balance}
\usepackage{comment}
\usepackage{enumerate}
\usepackage{caption}

\graphicspath{{./figs/}{./plots}{./images/}}

\newcommand{\eat}[1]{}

\begin{document}

\title{Shake-n-Shack: Enabling Secure Data Exchange Between Smart Wearables via Handshakes}

\author{\IEEEauthorblockN{Yiran Shen}
\IEEEauthorblockA{College of Computer Science \& Technology\\
Harbin Engineering University, China\\
shenyiran@hrbeu.edu.cn}
\and
\IEEEauthorblockN{Fengyuan Yang}
\IEEEauthorblockA{College of Computer Science \& Technology \\
Harbin Engineering University, China\\
yangfengyuan@hrbeu.edu.cn}
\and
\IEEEauthorblockN{Bowen Du}
\IEEEauthorblockA{Department of Computer Science\\
University of Warwick, UK\\
b.du@warwick.ac.uk
}
\and
\IEEEauthorblockN{Weitao Xu}
\IEEEauthorblockA{College of Computer Science \& \\Software Engineering\\
Shenzhen University, China\\
xuweitao005@gmail.com
}
\and
\IEEEauthorblockN{Chengwen Luo}
\IEEEauthorblockA{College of Computer Science \& \\Software Engineering\\
Shenzhen University, China\\
chengwen@szu.edu.cn
}
\and
\IEEEauthorblockN{Hongkai Wen*\thanks{*Corresponding Author}}
\IEEEauthorblockA{Department of Computer Science\\
University of Warwick, UK\\
hongkai.wen@dcs.warwick.ac.uk}

}

\maketitle

\begin{abstract}
Since ancient Greece, handshaking has been commonly practiced between two people as a friendly gesture to express trust and respect, or form a mutual agreement. In this paper, we show that such \emph{physical} contact can be used to bootstrap secure \emph{cyber} contact between the smart devices worn by users. The key observation is that during handshaking, although belonged to two different users, the two hands involved in the shaking events are often rigidly connected, and therefore exhibit very similar motion patterns. We propose a novel {\sl Shake-n-Shack} system, which harvests motion data during user handshaking from the wrist worn smart devices such as smartwatches or fitness bands, and exploits the matching motion patterns to generate symmetric keys on both parties. The generated keys can be then used to establish a secure communication channel for exchanging data between devices. This provides a much more natural and user-friendly alternative for many applications, e.g. exchanging/sharing contact details, friending on social networks, or even making payments, since it doesn't involve extra bespoke hardware, nor require the users to perform pre-defined gestures. We implement the proposed Shake-n-Shack system on off-the-shelf smartwatches, and extensive evaluation shows that it can reliably generate 128-bit symmetric keys just after around 1s of handshaking (with success rate $>$99\%), and is resilient to real-time mimicking attacks: in our experiments the Equal Error Rate (EER) is only 1.6\% on average. We also show that the proposed Shake-n-Shack system can be extremely lightweight, and is able to run in-situ on the resource-constrained smartwatches without incurring excessive resource consumption.
\end{abstract}
\IEEEpeerreviewmaketitle

\section{Introduction}
\label{sec:introduction}

\begin{figure}[htb]
\centering
\includegraphics[width=\columnwidth]{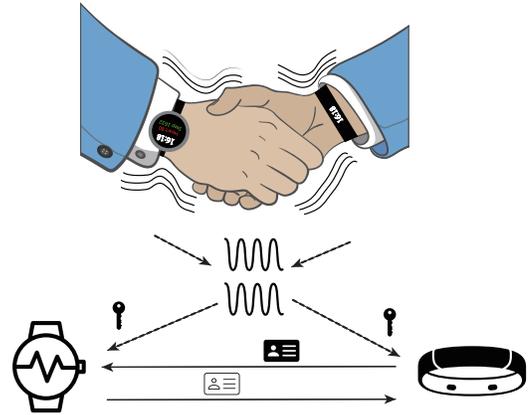}

\caption{The motion pattern induced by handshaking between two users can be captured by each of their wrist worn smart wearables (e.g. a smartwatch and a fitness band as shown in this figure), and used to generate cryptographic keys for secure data exchange, e.g. sharing contact details or friending on social networks.}
\label{fig:handshake}
\end{figure}

Wrist worn smart devices such as smartwatches and fitness bands are becoming ubiquitous: according to the latest global forecast~\cite{watch2020} their market is set to triple its volume in the near future, reaching \$32.9 billion by 2020. Instead of remaining as the companion devices of smartphones, now they are more independent, and capable of offering full-fledged functionalities. For instance, the recently announced Apple Watch Series 3 has the same level of connectivity including 4G LTE with smartphones, and can perform all kinds of tasks such as messaging, receiving/making calls, or streaming music without presence of the paired phone~\cite{apple2017independant}. Comparing to smartphones, those wrist worn wearables are often rigidly attached to the users' body, and equipped with rich sensing modalities, which makes them the ideal pervasive platform to sense and interact with the internal and external user states. In the near future they will become the key element in the cyber-physical ecosystem, enabling the next generation applications in a broad spectrum of sectors such as intelligent mobility, smart spaces, and digital healthcare~\cite{rallapalli2014enabling, atzori2010internet, mariakakis2016watchudrive, sen2016did}.

In this paper, we leverage the unique advantages offered by wrist worn smart devices, and explore the feasibility of using \emph{handshakes}, a common form of physical contact between human beings, to enable secure data exchange between their devices. The intuition is that in many circumstances, we often shake hands and then exchange physical or digital tokens in order to connect with the others. For example when meeting with someone for the first time, as a proper etiquette we typically shake hands with each other, and then exchange business cards, save contact details on our phones, or even friend each other on social networks. Essentially in those cases, handshakes can be viewed as the trigger for the subsequent data exchange activities, i.e. by shaking hands, both parties establish mutual consent to share information in physical or cyber world. Therefore, we aim to integrate the process of exchanging private information with the actual handshakes, so that when two users are shaking hands, their smartwatches or fitness bands can automatically communicate to each other and exchange data on-the-fly. We envision that in the future, users may have the option to configure their smartwatches or fitness bands to `socially discoverable' during handshakes, in the same way as the current wireless file sharing mechanisms such as the Apple AirDrop~\cite{wiki:AirDrop}.

Although this vision is appealing, there are several challenges need to be tackled. Firstly, the wireless communication channels established for data exchange have to be \emph{secured}, since information such as contact details is sensitive, and wireless communications are often prune to eavesdropping attacks. Secondly, the data exchange process should require \emph{zero effort} from the users, i.e. they only need to shake hands normally without extra intervention such as pairing in prior or entering the same PINs. Finally, the system should be efficient and lightweight enough to run \emph{in-situ} on the resource-constrained smartwatches or fitness bands, but not to quickly drain the device battery. 

Unfortunately, there is no existing solution can address all three challenges at the same time. For instance, some existing products and patents~\cite{razer,schorsch2013gesture} can detect handshakes or high fives to exchange data such as social media info, but they either broadcast over open wireless channels which can be easily intercepted~\cite{razer}, or rely on the cloud to verify keys~\cite{schorsch2013gesture}, which won't work without Internet connectivity and may incur undesirable delays. On the other hand, using key distribution protocols such as Diffie-Hellman (D-H) key exchange~\cite{diffie1976new} requires public key management infrastructure, which is computationally intensive and not feasible for real-time execution on wearables. The Near Field Communication (NFC) technology enables secure data communication between smart devices in the vicinity (normally within $20$cm). However, for now the security of NFC on smart wearables is largely guaranteed by the paired smartphones, such as passcodes or fingerprint authentication. As wearables can work independently of phone usage, the private data held by NFC can be vulnerable without the extra layer of protection.

To overcome the shortcomings of existing approaches, we propose the design and implementation of \textbf{Shake-n-Shack}, a novel system that directly uses handshakes to encrypt data exchange between wrist worn smart devices. It exploits the fact that within an episode of handshaking, the two hands holding together should produce similar movements, which can be picked up by sensors (e.g. accelerometers) of smartwatches or fitness bands worn on the corresponding wrists of both users. With the captured motion signals, the proposed Shake-n-Shack system generates symmetric cryptographic keys on both sides, which are used to secure subsequent data exchange between the two devices. Concretely, the technical contributions of this paper are as follows: 

\begin{itemize}
\item We propose Shake-n-Shack, a novel system for secure data exchange between wrist worn smart wearable devices via handshakes. To the best of our knowledge, this is the first work that explicitly uses physical contact (i.e. handshakes) to secure cyber contact (i.e. data exchange between smart devices) in a natural and user-friendly way.

\item We implement Shake-n-Shack on off-the-shelf smartwatches, and propose a set of efficient algorithms which capture and process the motion signals of handshaking events, and generate symmetric keys in a distributed fashion to encrypt/decrypt data exchange. Our system is lightweight and can be always-on: it runs in real-time on the resource-constrained smart wearables and incurs marginal overhead.

\item We evaluate Shake-n-Shack extensively on datasets collected from real-world settings. The results show that it is able to generate keys of over $140$ bits within 2s of handshaking, and the average key agreement rate between two legitimate devices are close to $100\%$. We also show that the proposed Shake-n-Shack system is resilient to run-time mimicking attacks, where the Equal Error Rate (EER) is only $1.6\%$. 

\end{itemize}

The rest of the paper is organized as follows. Sec.~\ref{sec:app_threat} discusses the application scenarios of the proposed Shake-n-Shack system, and describes the threat model. Sec.~\ref{sec:system_architecture} presents the overview of the Shake-n-Shack system, and Sec.~\ref{sec:algo} explains the details of the proposed approach that enables secure data exchange via handshakes. The proposed Shake-n-Shack system is evaluated in Sec.~\ref{sec:evaluation}, and related work is covered in Sec.~\ref{sec:related}. We conclude the paper in Sec.~\ref{sec:conclusion} and discuss potential future directions.

\section{Application Scenarios and Threat Model}
\label{sec:app_threat}

\begin{figure*}[t]
\centering
\includegraphics[width=\textwidth]{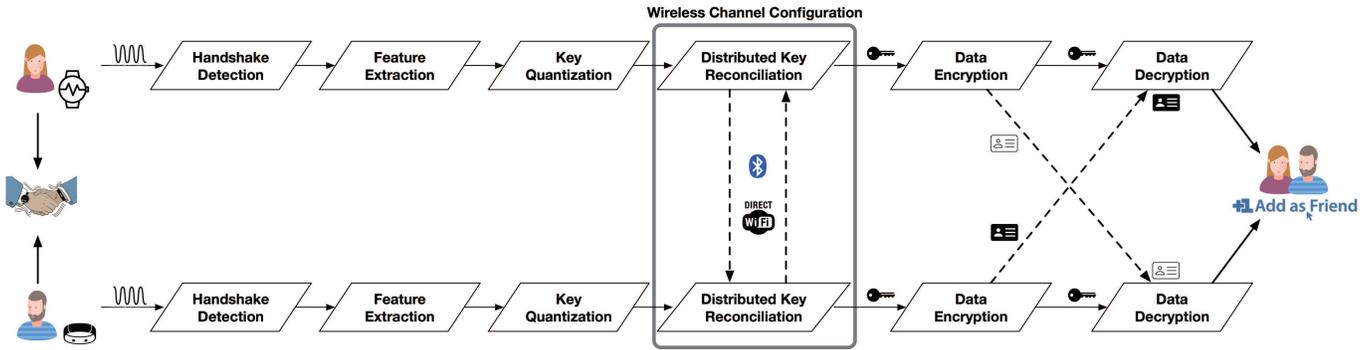}
\caption{The workflow of the proposed Shack-n-Shack system in the \emph{friending upon greeting} scenario as discussed in Sec.\ref{subsec:app}}
\label{fig:system_architecture}
\end{figure*}

\subsection{Application Scenarios}
\label{subsec:app}
The proposed Shake-n-Shack system provides a natural and reliable way to secure data exchange between wrist worn smart devices when shaking hands. It uses the motion signals induced by handshaking to simultaneously generating symmetric keys across different devices, and use the keys to establish secure communication channels. In fact, the generated keys can also be used to encrypt wireless communications between other mobile devices of the users, such as phones, laptops or even smart vehicles, as long as they remain paired or authenticated with the users' smartwatches or fitness bands. Therefore, the capability offered by Shake-n-Shack can be orthogonal to devices types or communication modalities, and is a fundamental building block for many application scenarios. In practice, it can be implemented as an OS level service, which empowers different  apps to exchange information securely with other users' devices. In the following, we discuss two example applications of the proposed Shake-n-Shack system.

\noindent \textbf{Friending upon Greeting: }
In many social events, connecting on social networks has become the convention after greeting with each other. With Shake-n-Shack, this process can be seamlessly done as two users shake hands, where they don't have to keep business cards or add each other manually on WeChat/WhatsApp. Concretely, during handshaking the data containing account details or friend request/confirmation will be encrypted at one of the two devices with the key generated by Shake-n-Shack, and transmitted through wireless channels. Then the other device uses the symmetric key to decrypt the data and perform further actions such as updating the friending status accordingly. 

\noindent \textbf{Instant Data Sharing: }
Shake-n-Shack can also be used to enable secure instant data sharing between users. As discussed above, other types of smart devices such as smartphones or tablets may also use the key generated by Shake-n-Shack to secure their communication channels. For instance, imagine Alice would like to share a photo on her laptop to Bob's phone. By simply shaking hands, her tablet uses the generated key to secure a wireless channel such as WiFi Direct, and transmits the data instantly. On Bob's side, he may use the symmetric key obtained from his smartwatch to decrypt and recover the photo, without even taking his phone out. 

\subsection{Treat Model}
Shake-n-Shack is designed to address \emph{impersonation attack}~\cite{ku2005impersonation}, which is a common security issue when transmitting data over wireless medium. In such attacks, an adversary impersonates to be legitimate in a communication protocol to extract private information exchanged by the legitimate devices. We consider the adversary as an active copycat, who observes and eavesdrops the data exchange activities among nearby legitimate users. As in~\cite{xu2017gait, mayrhofer2009shake}, in this paper we assume the adversary has full knowledge of the system details of Shake-n-Shack, and is able to sniff all the wireless traffic. However, we assume the legitimate devices have not been compromised, and it is impossible for the adversary to obtain the on-board motion data. Once the adversary extracts an encrypted message exchanged between two legitimate users, he tries to mimic the very handshake movements between them wearing a smartwatch or fitness band, by himself or with another adversary. The recorded motion data is then used to generate the same cryptographic key and attack the encrypted message. 

It is also worth pointing out that, in some cases the adversary may be able to record the handshakes using video cameras, and then try exhaustive search to decode the cached encrypted messages. However, in practice the cost of launching such sophisticated attacks is much higher, and can be potentially neutralized by making the data self-destruct~\cite{geambasu2009vanish} after a specific time. On the other hand, in some applications such as automatic friending on social networks, the exchanged data is merely short-lived tokens, which would already expire before the adversary could decode the messages.

\section{System Overview}
\label{sec:system_architecture}
In this section, we present an overview of the proposed Shake-n-Shack system. We consider the example of automatic friending on social networks as discussed in Sec.~\ref{subsec:app}, and show how Shake-n-Shack works in practice. Fig.~\ref{fig:system_architecture} demonstrates the workflow of the system in this scenario.

\noindent \textbf{Handshake Motion Capture: }
As two users start to shake hands, Shake-n-Shack captures the motion signals with the on-board Inertial Measurement Units (IMUs), which have been embedded in most of the current smart wearables. The segments of raw sensor data is firstly aligned and preprocessed, which are then used for the next step of key generation. 

\noindent \textbf{Key Generation \& Reconciliation: }
With the preprocessed motion data, Shake-n-Shack firstly applies dimensionality reduction techniques to recover the dominant signal components. The extracted signals are then quantized and converted into bits by thresholding. Due to the noisy nature of motion signals, Shake-n-Shack also incorporates a key reconciliation step with the other user's device, to discard the ambiguous bits and only keep the reliable ones to generate symmetric keys. 

\noindent \textbf{Wireless Channel Configuration: }
When a handshaking event is detected, Shake-n-Shack enables wireless radios on the smart wearables. As discussed in the previous section, Shake-n-Shack is communication modality neutral: it relies on the underlying network service to discover nearby devices, and tries to establish a wireless communication channel with the correct peer (i.e. the one party involved in this handshaking event) using the generated keys. 

\noindent \textbf{Secure Data Exchange: }
Once appropriate wireless communication channel is established, the proposed Shake-n-Shack uses the symmetric keys on both devices to secure data exchange between them. Concretely, we consider the standard messages encryption and decryption techniques, where the sender and receiver encrypts/decrypts the messages with their local keys respectively. In this way, although the wireless medium can be sniffed, the adversary won't be able to decode the intercepted data without the correct keys.

Now we are in a position to explain the details of the proposed secure data exchange approach.

\section{Secure Data Exchange via Handshakes}
\label{sec:algo}
In this section, we discuss the key components of the proposed Shake-n-Shack system which enable secure data exchange via handshakes. We start with how to detect and capture motion data during handshaking in Sec.~\ref{subsec:handshake_detection}, and then describe our algorithms to generate and reconcile keys in Sec.~\ref{subsec:key-gen} and Sec.~\ref{subsec:key-rec} respectively. Then in Sec.~\ref{subsec:channel-est} we show how to establish the correct wireless channel, and finally Sec.~\ref{subsec:enc_dec} discusses the process of data encryption \& decryption in the proposed Shake-n-Shack system.

\subsection{Handshake Motion Capture}
\label{subsec:handshake_detection}
In this paper, we assume that both users involved in a handshaking event wear smart devices such as smartwatches or fitness bands, on the wrists of their dominant hands, i.e. the hands that are used during the handshakes. In this case, the motion of the handshakes can be recorded by the Inertial Measurement Unit (IMU) sensors, which are embedded in most of the current off-the-shelf smart wearables. Typically, they contain a 3-axes accelerometer, and optionally gyroscope and magnetometer. In our system, we only use accelerometers since they are the most pervasive and very efficient, while gyroscope consumes significantly more energy. Our experiments show that, with sensor readings just from the accelerometers, the proposed system is able to generate robust enough cryptographic keys.

\begin{figure}
\centering
\includegraphics[width=\columnwidth]{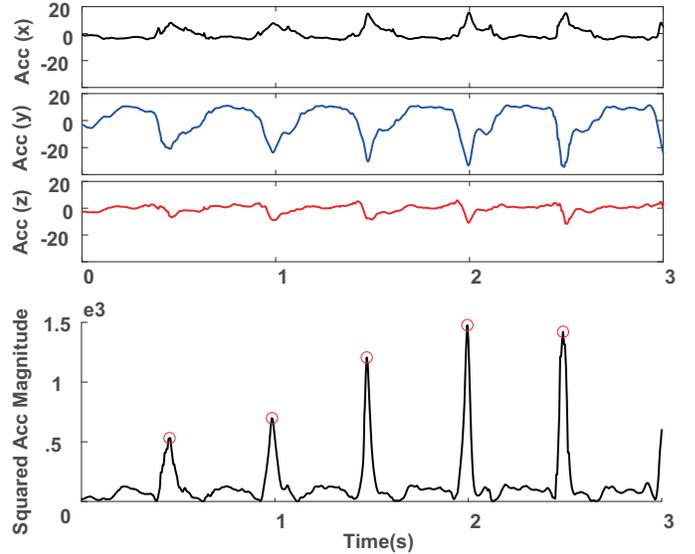}
\caption{The acceleration readings along three axis during 3s of a handshaking event (top three plots), and the squared acceleration magnitude (the bottom plot). }
\label{fig:handshake_detection}
\end{figure}

Fig.~\ref{fig:handshake_detection} shows an example of the accelerometer sensor readings over three axes during a handshaking event (top three plots), where we can see that the handshake induces periodic patterns in the signal. On the other hand, the bottom plot of Fig.~\ref{fig:handshake_detection} shows the squared magnitude of acceleration, which is computed by combining the squared data values of signals from all three axes. Clearly, the acceleration magnitude is very significant with respect to the background noise, and we see several peaks corresponding to the up and down movement during handshaking. Therefore, our system detects the first signal peak in a handshake event, and use it as the anchor point to align the sensor readings for the next step of key generation.  


\subsection{Key Generation}
\label{subsec:key-gen}
\noindent \textbf{Signal Feature Extraction: }
As shown in Fig.~\ref{fig:handshake_detection}, the raw acceleration signal has three axes, representing motion with respect to the body reference frame of the device. In practice, the devices can be attached to the wrists of the users in arbitrary ways, to reliably generate symmetric keys across different devices, we need to transform the signal to reveal the dominant motion patters that are invariant to device attachment. On the other hand, although the acceleration magnitude is robust to different device posture, as shown later in Sec.~\ref{subsec:eval-setup} it contains much less information and thus impossible to generate high quality keys (i.e. in terms of bit rate). Therefore in this paper, we consider a Principal Component Analysis (PCA) based approach, which converts the 3-axes acceleration into signals representing the dominant motion components. 

\begin{figure}[t]
\centering
\includegraphics[width=\columnwidth]{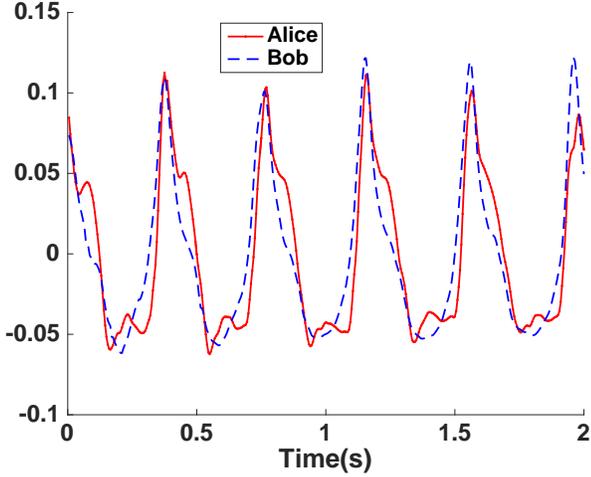}
\caption{The proposed PCA-based approach extracts dominant motion features from raw acceleration data, which are very consistent across the two users shaking hands with each other (Alice and Bob in this case).}
\label{fig:pca}
\end{figure}

Let us assume two users, \emph{Alice} and \emph{Bob}, have shaken hands with each other, and the motion data has been captured by both of their smart wearables. Suppose Alice obtains a data matrix $X\in\mathcal{R}^{M\times N}$ containing the accelerometer readings, where $M$ is the number of axes and $N$ is the number of data points from each axis. PCA finds a matrix $U\in\mathcal{R}^{m\times M}$ which projects the original data into a smaller subspace while retaining most of the information. Before computing the matrix $U$, the data matrix should be firstly centered., i.e., the mean value of each column in data matrix $X$ is subtracted:

\begin{equation}
\label{eq:quantization1}
X_{C} = X - \frac{1}{m}\sum_{i=1}^MX_i 
\end{equation} 
where $X_i$ is the $i_{th}$ row of $X$. After the data matrix is centered, the Eigenvalues and Eigenvectors matrices can be obtained by conducting EigenValue Decomposition (EVD) as,
\begin{equation}
\label{eq:quantization2}
\{ U, E_{ordered}, V^T \} = EVD(X_CX_C^T)
\end{equation}
$E_{ordered}$ contains the Eigenvalues on its diagonal, which are sorted in descending order according to their absolute values. $U$ contains the corresponding Eigenvectors by columns. Then we project the original data matrix into the subspace defined by the $U$:
\begin{equation}
Y = U^TX 
\end{equation}
where $Y\in\mathcal{R}^{m\times N}$.  It is well known that the first principal component contains the largest part of information in the original signal. It corresponds to the largest Eigenvalue in $E_{ordered}$, and the first row of the matrix $Y$. Therefore, we use the first row of $Y$ (denoted as $Y^1$) to represent the dominant motion signal features. Fig~\ref{fig:pca} shows an example of the computed features for both Alice and Bob. We see that although generated on different devices, they exhibit similar patterns and agree to each other very well. This confirms that it is clearly possible to use such features to generate symmetric keys across two different devices.

\noindent \textbf{Signal Quantization: }
Now we need to convert the continuous motion signal features into discrete keys. Our system uses a similar signal quantization and bit extraction methods as in~\cite{javali2015secret} and~\cite{revadigar2015dlink}, which quantize continuous signals into bit codes. The idea is to firstly divide the time-varying continuous signals into small segments with the same length. e.g. 10 data points as in our experiments. Then for each segment, we compute its upper ($\delta_+$) and lower ($\delta_-$) quantization thresholds as:
\begin{equation}
\begin{array}{l}
\delta_+= \mu + K\sigma\\
\delta_- = \mu-K\sigma
\end{array}
\end{equation}
where $\mu$ is the mean and $\sigma$ is the standard deviation of the values of the data points within the segment. $K$ is the quantization factor, which determines the bit generation rate and key agreement rate (we will discuss this in more details in Sec.~\ref{subsec:eval-setup}).

In practice, those thresholds are used to determine the bit values at each position of the generated key. Specifically, a data point over $\delta_+$ will be encoded as $1$, while the one below $\delta_-$ will be $0$. Any data points between the thresholds will be discarded to improve stability. After this quantization step, now Alice and Bob have their own keys locally at their devices. Without loss of generality, in the following we denoted the keys as $\mathcal{K}_{Alice}$ and $\mathcal{K}_{Bob}$.

\subsection{Key Reconciliation}
\label{subsec:key-rec}
Theoretically, the generated keys can be used to encrypt and decrypt the messages if and only if they agree with each other, i.e. $\mathcal{K}_{Alice}$ = $\mathcal{K}_{Bob}$. However in practice, the keys generated from the above process may not be able to precisely agree with each other due to signal noise. In most cases, we would obtain two keys $\mathcal{K}_{Alice}$ and $\mathcal{K}_{Bob}$, where $\mathcal{K}_{Alice}\approx\mathcal{K}_{Bob}$. The disagreement often happens if the value of a data point is close to the thresholds, i.e. in the presence of small turbulence caused by motion noise, a bit at Alice's side might be discarded since the value is below the upper threshold $\delta_+$, while remained `1' at Bob's side. 

To address this problem, we consider a key reconciliation approach~\cite{xu2017gait} to discard those ambiguous bits. The idea is to exchange the index of the valid bit positions to the other devices, and reach a mutual agreement on which bits should be used in the final keys. For instance, assume the key generated by Alice's devices is [1x0xx11x00], while for Bob is [1x00x11xx0], here x means the position where no valid bit is present. Then both Alice and Bob inform each other the positions of the valid bits, i.e., Alice sends $P_{Alice}=\{1, 3, 6, 7, 9, 10\}$, and Bob sends $P_{Bob}=\{1, 3, 4, 6, 7, 10\}$. Upon receiving the positions, they compare the received vector with the local one, and agree that only the bits that are valid according to both vectors should be used. In this example, the agreed positions should be $\{1, 3, 6, 7 , 10\}$ so that the final symmetric keys are $\mathcal{K}_{Alice}$ = $\mathcal{K}_{Bob}$ = [10110].

Note that at this stage, it is possible that the two users haven't established a communication channel with each other. They just broadcast and receive the position vectors of valid bits to/from the nearby devices. For instance, there may be two pairs of users using Shake-n-Shack in a close proximity, and in the following we explain how Shake-n-Shack uses the received information to establish a secure wireless channel with the correct peer.   

\subsection{Wireless Channel Configuration}
\label{subsec:channel-est}
As discussed above, in the presence of multiple pairs of Shake-n-Shack users, we need to correctly set up communication channels between the right peers, i.e. the two who are shaking each other's hands. One naive approach is to look at the received signal strengths (RSS), and choose the device with the strongest RSS to connect. However this is not robust enough in practice since RSS measurements are inherently noisy. The proposed Shake-n-Shack considers a probe-based approach. For instance, Alice might receive more than one position vectors $P$ from different devices, including Bob's. Then it performs the above reconciliation step for each vector, and generate a list of candidate keys. Those candidate keys are then used to encrypt a pre-defined probe message whose content is known by everyone, and the encrypted messages (one for each candidate key) are replied back to the corresponding senders. 

On the other hand, When Bob receives those probe messages, it tries to use the list of candidate keys to decrypt the messages accordingly. It should be that just one message can be successfully decrypted, which is the one sent by Alice. Now Bob only needs to keep the candidate key from Alice, and reply an acknowledgement message to Alice, indicating that he is ready for further data exchange. In this way, the proposed Shake-n-Shack system establishes a wireless communication channel between Alice and Bob who is shaking hands with each other, and identifies the correct keys for future data encryption and decryption.

Note that besides the two wearables that have generated the keys (i.e. the smartwatches or fitness bands worn by the users), such a communication channel could also be established between other devices (e.g. phones or laptops) belonging to the users, as long as they are paired or connected with those wearables in the same private (often ad hoc) network. This is straightforward since the wearables can simply share the generated keys with the other devices within the network, allowing those devices to directly perform the above probe process, and initialize data exchange with each other. 

\subsection{Secure Data Exchange}
\label{subsec:enc_dec}
Given the symmetric keys and configured wireless communication channel, the proposed Shake-n-Shack uses the standard encryption and decryption methods to guarantee the confidentiality of the data exchange. In our implementation, we require the generated keys to contain at least 140 valid bits to ensure the level of security, and if unsuccessful Shake-n-Shack will ask the users to do another handshake (see {\sl Resilience to Mimicking Attacks} in Sec.~\ref{subsec:eval-setup}). Then we consider a fixed length of the first 128 bits to encrypt and decrypt messages at both sides. 

Depending on different application scenarios, in some cases such as automatic friending on social networks, the data exchanged is merely digital tokens, i.e. friend requests/confirmations. Therefore after successful completion of the data exchange, the proposed system will also inform the social networks about this transaction via their APIs, to update the friending status accordingly. 

\section{Evaluation}
\label{sec:evaluation}

\subsection{Experiment Setup}
\label{subsec:eval-setup}

\begin{figure*}[!tbp]
  \centering
  \begin{minipage}[b]{0.35\textwidth}
    \includegraphics[width=\textwidth]{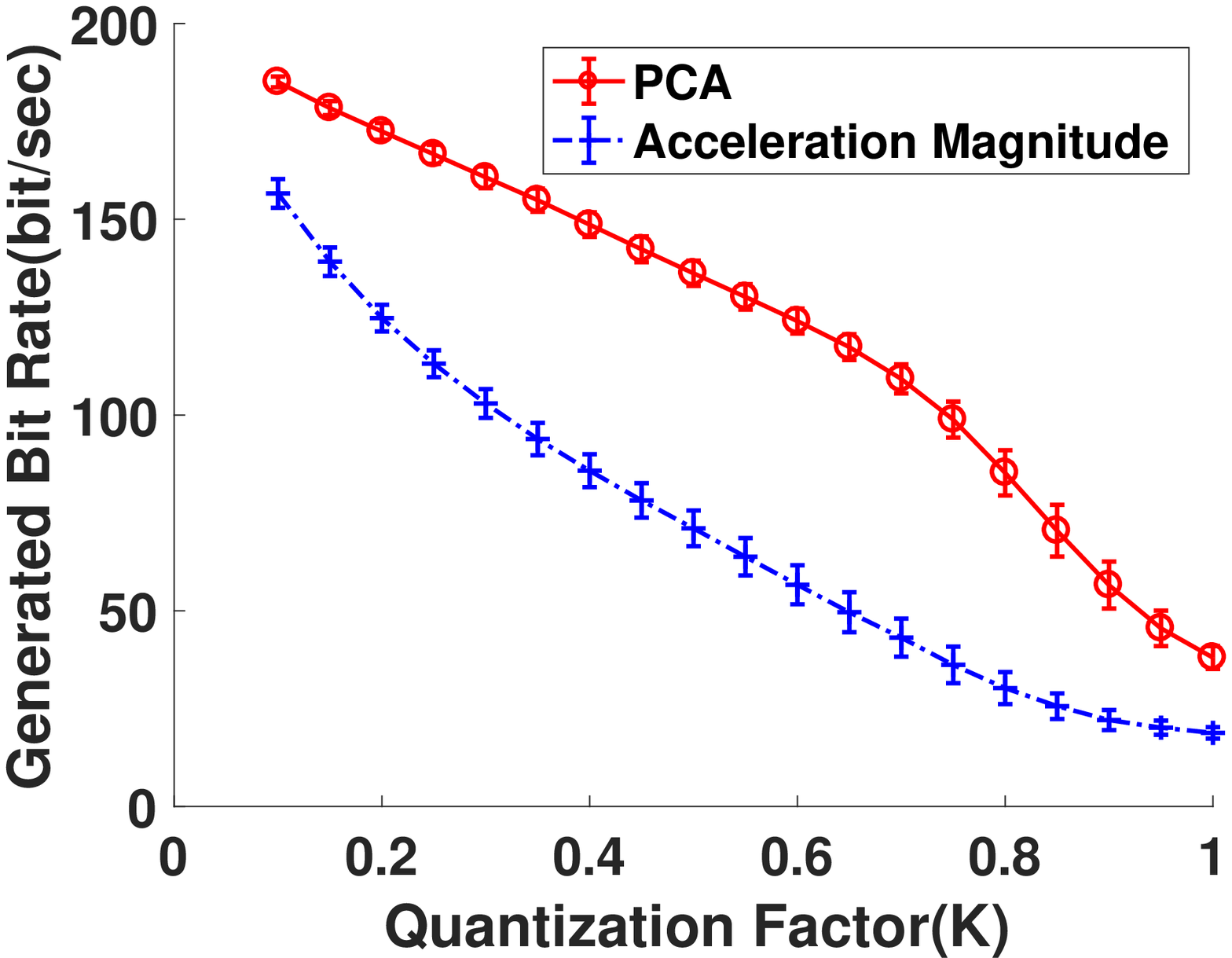}
    \caption{The bit rate of keys under different quantisation factor $K$, generated by the proposed PCA-based approach (the red line) and the baseline approach using acceleration magnitude (the blue line).}
    \label{fig:pca_vs_power}
  \end{minipage}
  \hfill
  \begin{minipage}[b]{0.63\textwidth}
    \includegraphics[width=\textwidth]{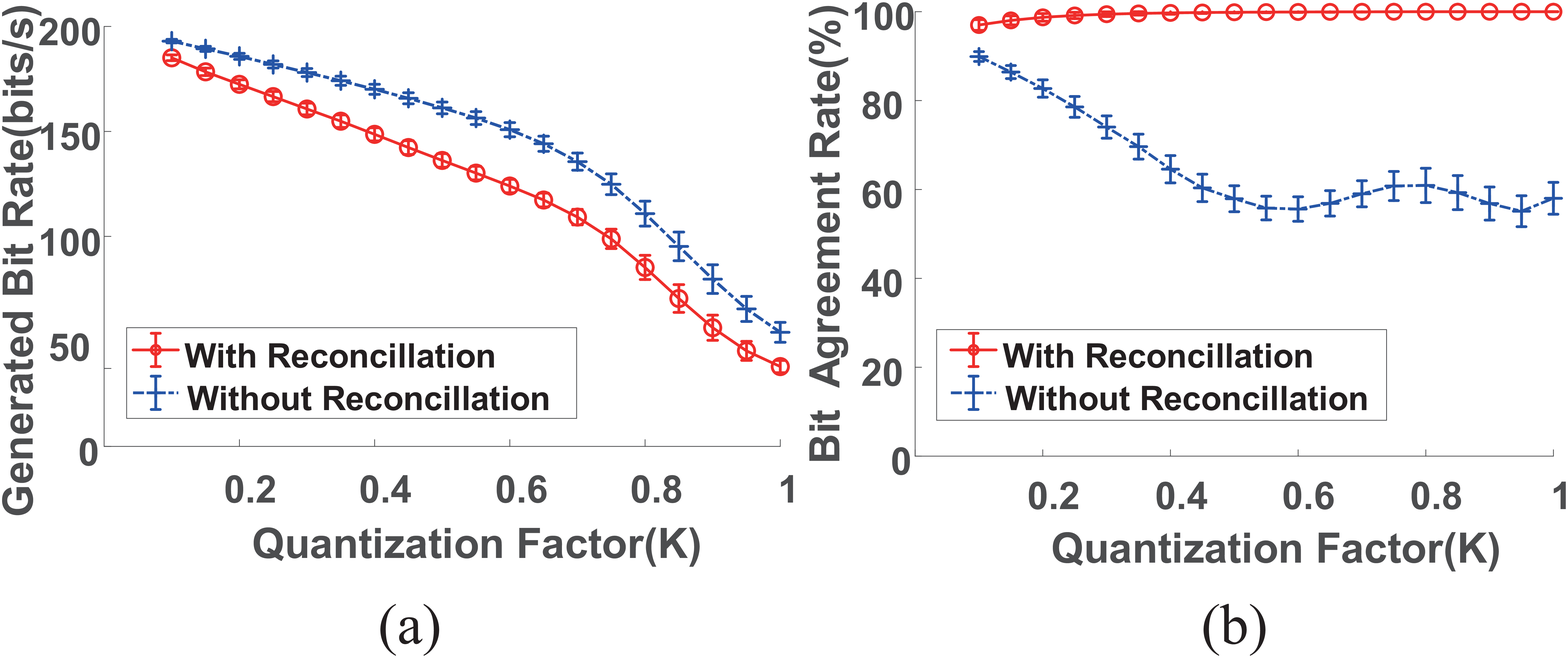}
    \caption{The impact of key reconciliation on (a) bit rate, and (b) bit agreement rate of the generated keys.}
    \label{fig:reconciliation}
  \end{minipage}
\end{figure*}

\noindent \textbf{System Implementation: }
We implement Shake-n-Shack on off-the-shelf smartwatches, which can run in real-time. In our experiments, we use Samsung Gear Live, which has a Quad-core 1.4GHz CPU, 512M RAM, and 300mAh battery, and run Android Wear OS. We set the accelerometer sampling rate to 200Hz, and uses the Bluetooth 4.0+ {\sl InsecureRfcomm} to establish wireless communication channels.

\noindent \textbf{Data Collection: }
We recruited $20$ volunteers to participate in our experiments, containing $10$ males and $10$ females with age ranging from $22$ to $54$. During each experiment session, we asked all participants to wear smartwatches on their right wrists, and randomly divided the $20$ participants into five groups, each containing four people. Within each group, two of the participants were selected to be the legitimate users while the other two were the adversaries. Then the legitimate users were asked to shake hands to exchange data, while at the same time the adversaries observed this and also shake hands with each other, trying to mimic the pattern of the handshake as much as they can.

We conduced over $1, 000$ sessions of experiment for two weeks time, and have collected two datasets: one containing motion data from the legitimate users, while the other is the data from the adversaries. In the following unless otherwise stated, we refer to those two datasets as the \emph{legitimate dataset} and \emph{adversarial dataset}.

\noindent \textbf{Evaluation Metrics: }
In this paper, we evaluate the performance of the proposed Shake-n-Shack system with respect to the following metrics. 
\emph{ Generated Bit Rate: } is the number of bits generated from the sensor readings per second. 
\emph{ Bit Agreement Rate} denotes the percentage of the matching bits of the two cryptographic keys generated by two devices during a handshaking event. 
\emph{ Signals Coherence} is the empirical CDF of the coherence of different motion signal features. 
\emph{ Key Success rate} is the percentage that the two keys generated via handshakes are identical. 
\emph{ False Acceptance Rate (FAR)} is a measure of the probability that a mimicking adversary generates an identical key to that of a legitimate device. 
\emph{ False Rejection Rate (FRR)} is defined as the ratio of the failed matching attempts via handshakes. It is obvious there is a trade-off between FAR and FRR. 
\emph{ Equal Error Rate (EER)} measures the trade-off between FAR and FRR and it is the value of FAR or FRR when the two false rates are equal.  
We also evaluate the system overhead of Shake-n-Shack by profiling the \emph{Computation Time} and \emph{Energy Consumption} of its key components on off-the-shelf smartwatches.

\subsection{Experiment Results}
\label{subsec:eval-setup}


\noindent \textbf{Bit Rate vs. Signal Feature Extraction: }
The first experiment studies the impact of signal feature extraction approaches in terms of the generated bit rate. Fig.~\ref{fig:pca_vs_power} shows the mean and standard deviation of the generated bit rates of the proposed PCA based approach with respect to the baseline method of using acceleration magnitude. First we see that the proposed approach produces significantly higher bit rate than using acceleration magnitude. For example, when $K=0.7$, the average generated bit rate from our approach is about $120$ bit/sec, which almost triples that of the acceleration magnitude based. In addition, we also see that as quantization factor goes up, the generated bit rate drops for both approaches. This is expected since high quantization factor means bigger thresholds, where the bits generation process is more prune to noise. However, the PCA based approach degrades much more gracefully than the baseline, especially in the range of $[0.4, 0.8]$. This is because our approach can recover the most dominant motion patterns from the raw signal, and thus is inherently more robust to high quantization factors.  



\noindent \textbf{Impact of Key Reconciliation: }
This experiment verifies the impact of the proposed key reconciliation method as discussed in Sec.~\ref{subsec:key-rec}. We consider both the generated bit rate and bit agreement rate as the metrics. As shown in Fig.~\ref{fig:reconciliation}(a), we see that on one hand, reconciliation slightly reduces the generated bit rate, since it discards those ambiguous bits. However, this has a positive knock-on effect on the bit agreement rate. As shown in Fig.~\ref{fig:reconciliation}(b), the bit agreement rate is significantly higher when key reconciliation is considered. In practice, for a reasonable $K$, we would prefer the high bit agreement rate introduced by reconciliation, since the generated keys are only valid if they are identical in every bit: if an agreement can't be reached, the users would have to shake hands again to regenerate the keys. In addition, the gap between with and without reconciliation in Fig.~\ref{fig:reconciliation}(a) is only marginal especially when $K<0.4$ and $K>0.7$. 

\begin{figure*}
\centering
\subfigure[Bit Rate]{
\epsfig{file=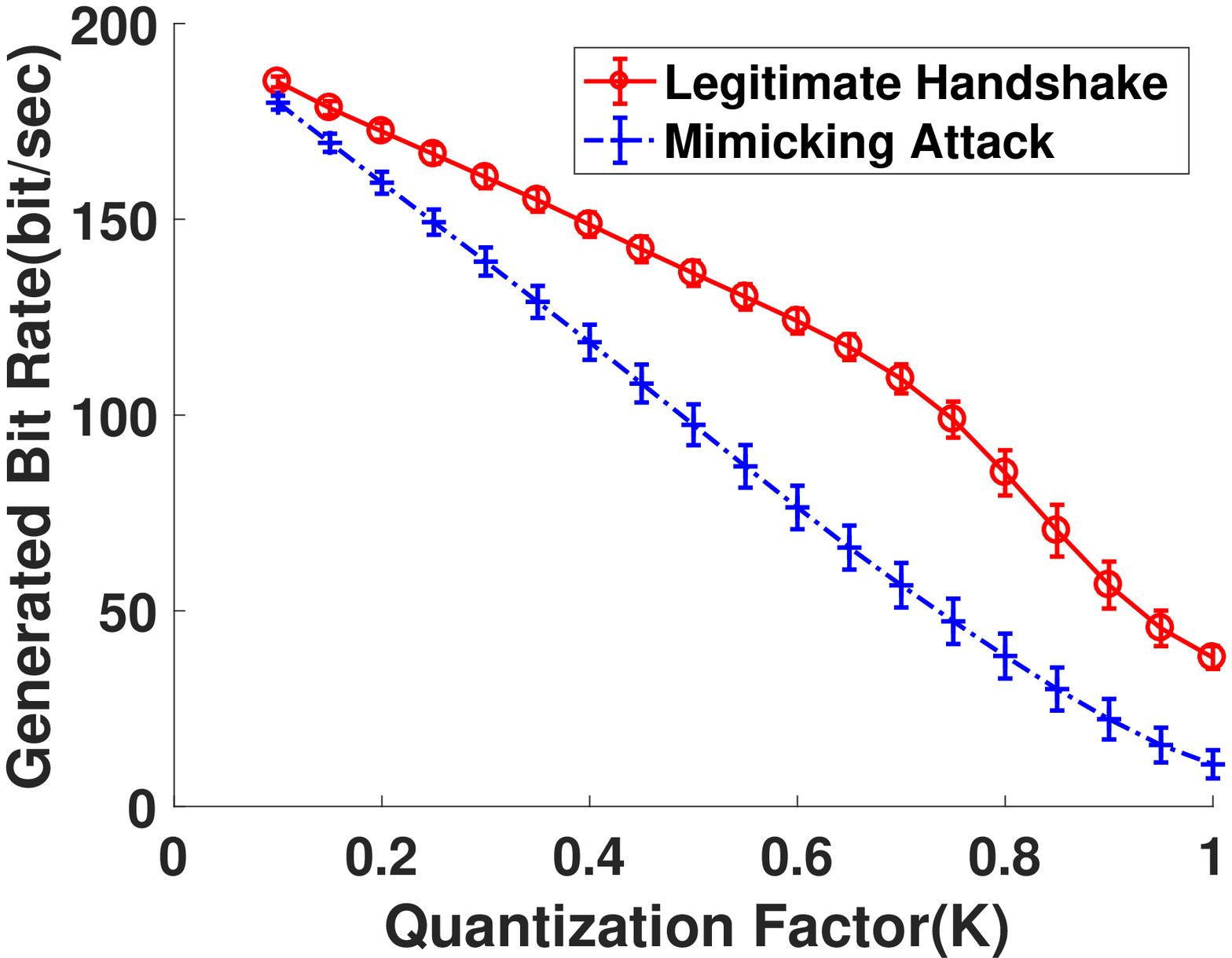,width = 2.2in}
\label{fig:k_bitsrate}
}
\subfigure[Bit Agreement Rate]{
\epsfig{file=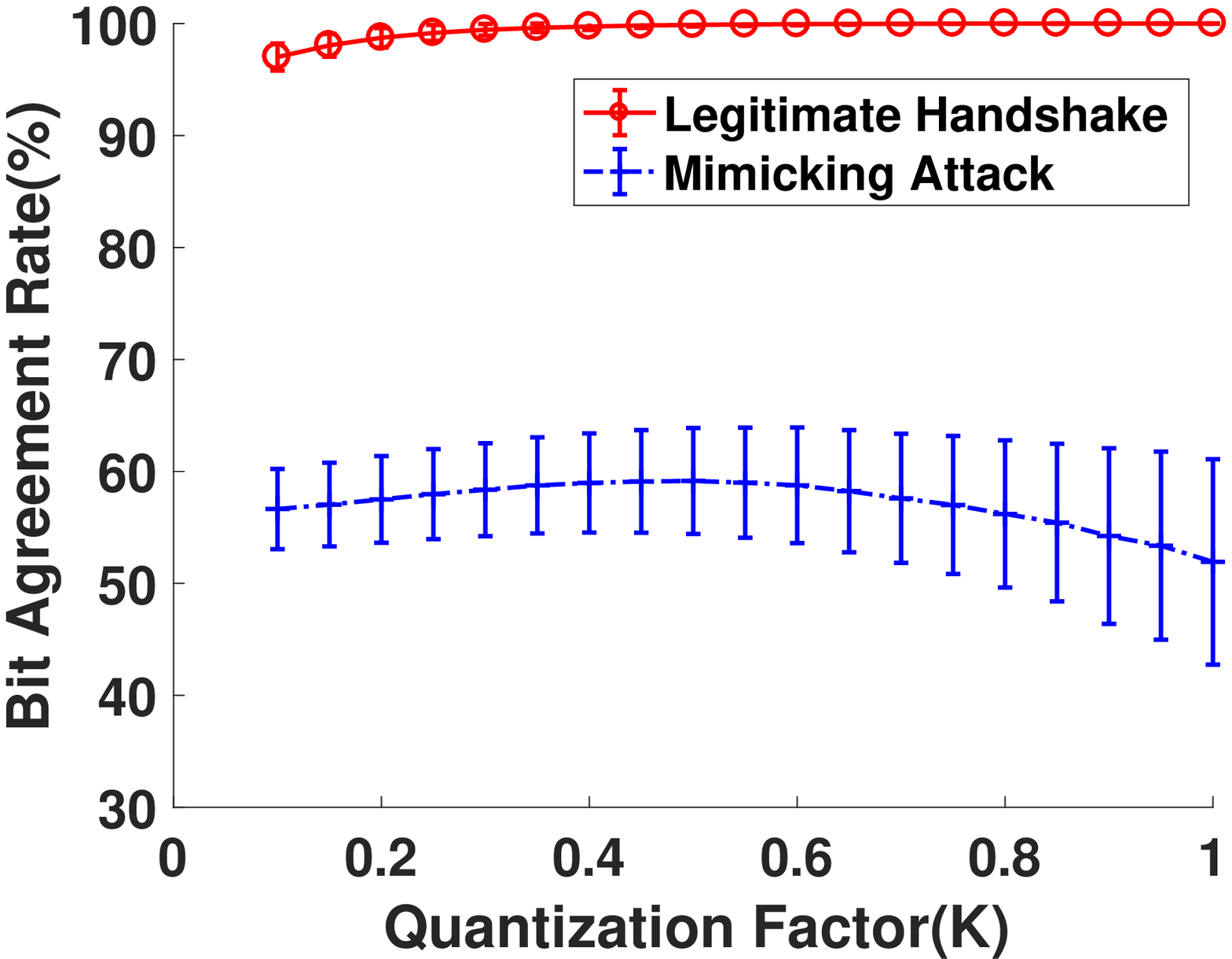, width=2.2in}
\label{fig:k_agree}
}
\subfigure[Key Success Rate]{
\epsfig{file=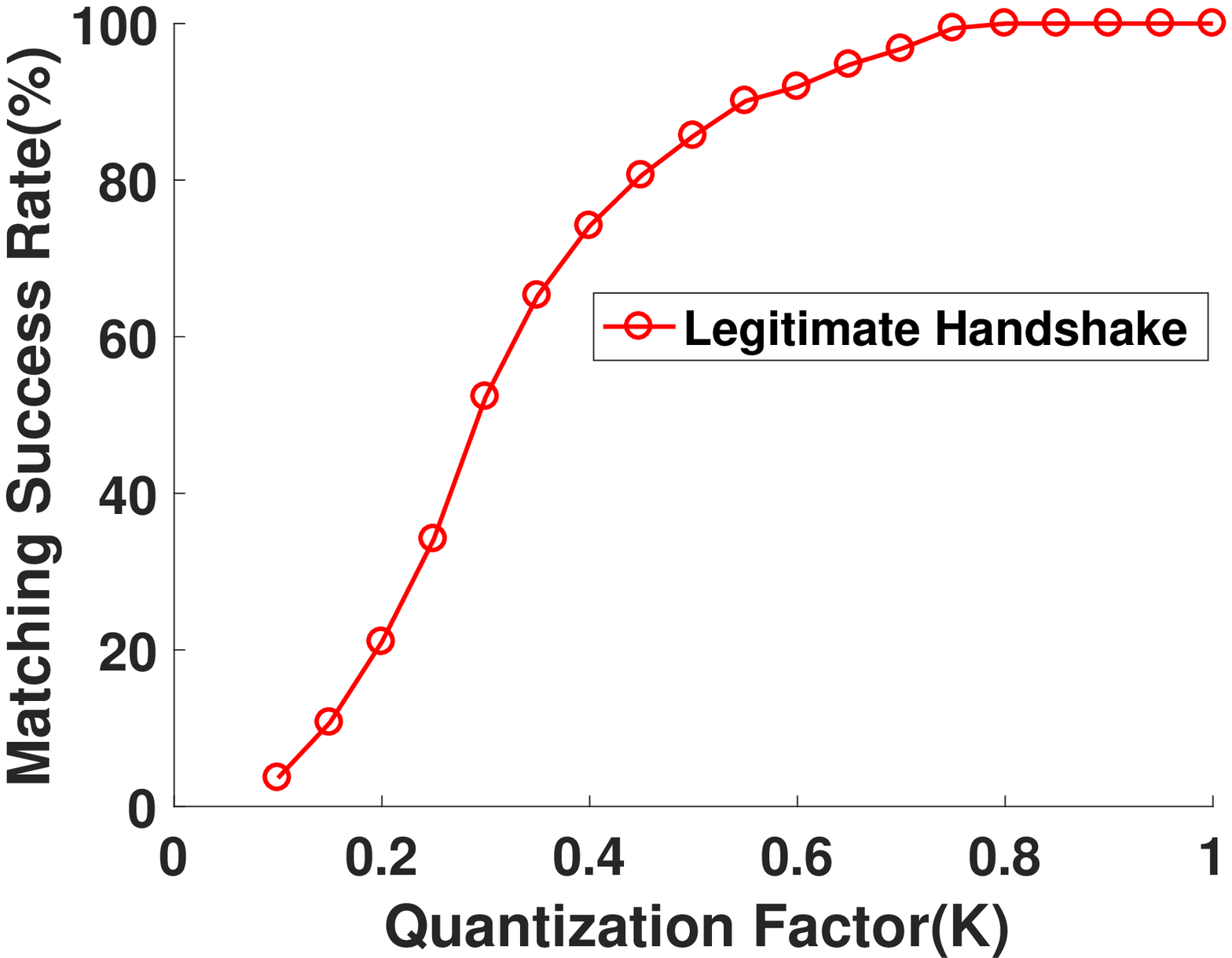, width=2.2in}
\label{fig:k_success}}
\caption{The impact of different quantization factor ($K$) on the system performance in terms of (a) bit rate, (b) bit agreement rate and (c) key success rate. }
\end{figure*}

\noindent \textbf{Sensitivity Analysis: }
This set of experiments aims to systematically investigate the sensitivity of the proposed Shake-n-Shack system. We use both the legitimate and adversarial datasets, and vary the quantization factor $K$. The following three metrics, generated bit rate, bit agreement rate and key success rate are considered in the experiments. 

Fig.~\ref{fig:k_bitsrate} shows how the generated bit rate varies over $K$ on both datasets. We see that generally a smaller $K$ produces higher bit rate, and handshakes made by legitimate users (i.e. from the legitimate dataset) achieve higher bit rate than that of the adversaries. This is because legitimate handshakes tend to produce very similar motion signals across two devices, and thus only fewer bits in the generated keys will be thrown away during the key reconciliation step. 

On the other hand, for bit agreement rate, as shown in Fig.~\ref{fig:k_agree}, legitimate handshakes have slightly higher bit agreement rate as $K$ goes up. However, for the handshakes produced by adversaries, the bit agreement rate first go slightly up and then dropped, and the variances becomes much bigger. Note that the average bit agreement rate from legitimate handshakes achieve almost $100\%$ when $K$ is larger than $0.7$. More importantly, we see that the gap between legitimate and adversarial handshakes is very significant for all $K$. This implies that legitimate handshakes are very hard to be mimicked in real-time, since it is very difficult for the adversaries to reliably generate high quality keys.

Finally, we evaluate how key success rate on legitimate dataset, i.e. the generated two keys are identical, varies with respect to the quantization factor $K$. As shown in Fig.~\ref{fig:k_success}, the key success rate increases as $K$, which reaches $100\%$ when $K$ is larger than $0.7$. Therefore, in our case a larger quantization factor $K$ results in higher bit agreement rate and key success rate, but has a negative impact on the bit rate. In addition, we observe that the bit rate and bit agreement rate of legitimate handshakes are consistently higher than that of the adversarial, which means we could reject the adversarial attempts with appropriate thresholds. Particularly in the current Shake-n-Shack system, we use bit rate threshold to decide if the generated key should be accepted as valid.

\begin{figure}[htb]
\centering
\epsfig{file=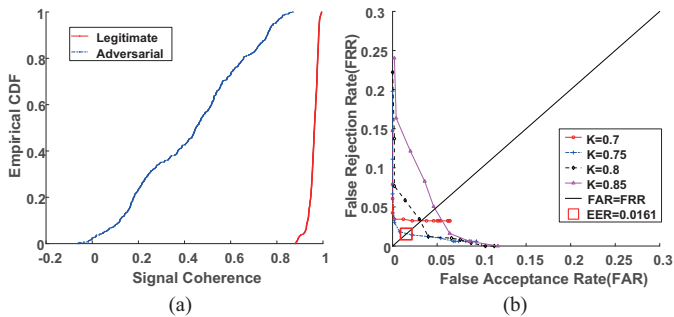, width=\columnwidth}
\caption{(a) The CDF of signal coherence between two signals generated by legitimate handshakes (red), and by an adversary mimicking the legitimate handshakes (blue). (b) FAR and FFR of Shake-n-Shack under different bit rate thresholds and quantization factor $K$. The best EER can be achieved in our experiments is $0.016$ (marked by the red rectangle).}
\label{fig:security}
\end{figure}

\noindent \textbf{Resilience to Mimicking Attacks: }
In this experiment we evaluate the performance of the proposed Shake-n-Shack system under mimicking attacks. We assume that as two legitimate users are shaking hands, there is an adversary who can sniff the wireless media, and mimic the handshaking patterns in real-time, trying to produce the same cryptographic key. For the handshaking events, we first analyze the coherence between: a) signals produced by two legitimate devices, and b) signals produced by one legitimate device and the adversarial device which was imitating that handshake. Fig.~\ref{fig:security}(a) shows the empirical CDFs of signal coherence for the two cases. We see that the legitimate handshakes consistently produce much coherent signals: around $97\%$ of the signals have coherence values over $0.9$, while for adversarial around $97\%$ is under $0.8$. This confirms that: a) during handshaking the two legitimate devices tend to induce very similar motion signals; and b) it is difficult to mimic the patterns of the handshakes in real-time even the adversaries are close by.

%

Based on the above observations, now we evaluate the performance of Shake-n-Shack when facing mimicking attacks using metrics \emph{ False Acceptance Rate (FAR)}, 
\emph{ False Rejection Rate (FRR)} and \emph{ Equal Error Rate (EER)}. Generally, we would like to reduce both FAR and FRR, but in practice they are often negatively correlated. As discussed in the previous experiments, we use bit rate threshold to determine if a key should be accepted. Therefore, here we vary the threshold (from $20$ to $90$) to see the correlation between FRR and FAR (see the lines in Fig.~\ref{fig:security}(b), where each dot is corresponding to a bit rate threshold value).  

We also vary the quantization factor $K$, and plot the FAR vs. FRR curves for $K=\{0.7, 0.75, 0.8, 0.85\}$. As we can see, Shake-n-Shack works best with $K=0.75$, where both FAR and FRR are lower (the curve is closer to the origin). In this case the EER is $0.016$ (see the rectangle marked in Fig.~\ref{fig:security}), which is the best balance between FAR and FRR. However in practice, we typically require a lower FAR to make the system more conservative and robust against mimicking attacks, although this will increase the chances of falsely rejected legitimate handshakes. Concretely, in the current implementation we set the quantization factor $K$ to $0.75$, and the bit rate threshold to $70$ bits per second. As shown in Fig.~\ref{fig:k_bitsrate}, with these parameter settings Shake-n-Shack is able to generate a 128-bit key with only on average $1.3$ seconds of handshaking, and we use device vibration patterns to inform the users if the data exchange is successfully completed, or requires another go.

\noindent \textbf{System Overhead: }
The final set of experiments evaluate the overhead of the proposed Shake-n-Shack system when running on off-the-shelf smart wearables. We profile the running time and energy consumption caused by key components of the system, including \emph{Handshake Detection}, \emph{Feature Extraction}, and \emph{Key Generation}. The overhead of rest components such as data encryption and decryption can be application-specific, and is not considered in this paper. In our implementation, we use the very lightweight Advanced Encryption Standard (ASE) for data encryption/decryption, which only incurs negligible overhead according to our previous work~\cite{xu2017gait}.

\begin{table}[t]
\centering
\caption{Overhead of the proposed Shake-n-Shack system inters of running time (ms) and energy consumption (mJ). }

\label{tab:resource_consumption}

\begin{tabular}{|c|c|c|c|}
\hline
Components&Time&Energy\\
\hline
Handshake Detection& 3.1 {\sl ms} & 2.04 {\sl mJ}\\
\hline
Feature Extraction& 78.4 {\sl ms}&60.35 {\sl mJ}\\
\hline
Key Generation& 89.6 {\sl ms}&6.25 {\sl mJ}\\
\hline
Total& 171.1 {\sl ms}& 68.64 {\sl mJ}\\
\hline
\end{tabular}

\end{table}

Table~\ref{tab:resource_consumption} presents the detailed resource consumptions of the three components obtained from Android APIs, averaging from $30$ independent tests. We can see that the total running time is only 171.1ms, and thus we could easily run Shake-n-Shack in real-time. On the other hand, the average extra energy consumed by Shake-n-Shack is 68.64 {\sl mJ}. Since Shake-n-Shack only works opportunistically when data exchange is required, its impact on the battery life should be limited. In fact, the battery capacity of Samsung Gear Live used in our experiments is $300$ mAh (i.e., 4.32 {\sl kJ}). If the targeted lifespan of the devices is one day (normally $12$ hours of active usage), the proposed Shake-n-Shack system only accounts for $0.019\%$ of the hourly budget ($360$ {\sl J}), which is marginal. 
\section{Related Work}
\label{sec:related}

\noindent \textbf{Wrist Worn Smart Wearables: }
The studies on smartwatches are booming in recent years and have produced many novel applications to improve the well-being of human's life. The sensor readings, e.g., accelerometer and gyroscope of the smartwatches can easily track the motion of user's arm therefore are often used in activities recognition. For examples, in~\cite{sen2016did}, the authors proposed to use smartwatches to detect if the wearers took a foosball break at workplace to maintain the work-break balance of the workers. Moreover, smartwatches were used to detect activities of the drivers in~\cite{mariakakis2016watchudrive} to ensure the safety of the wearer during drive and they could be also used to recognize the text inputs~\cite{arduser2016recognizing}. To improve the accuracy of activities recognition on smartwatches, the authors in~\cite{bhattacharya2016smart} sought the power of deep learning techniques. Besides of activities recognition, there is also some work concerning the security issue related to smartwatches. For example, in~\cite{xu2017gaitwatch} the authors proposed, Gait-watch, an authentication system exploiting the unique gait information of the smartwatch users to automatically detect the real user or malicious intruders. 

\noindent \textbf{Secure Wireless Communication Between Smart Devices: }
A secure communication channel establishment between smart mobile devices requires multiple parties encryption key generation. Key generation methods are mainly based on the common information between the two parties sharing data. One of the mainstream studies focus on the properties of the wireless communication. For example, the Received Signal Strength Indicator (RSSI) of the wireless channel are frequently used to produce shared keys~\cite{revadigar2015mobility, revadigar2015dlink, revadigar2016secure, javali2014seak, shi2013ask} as the RSSI at the both ends should be the same. However the RSSI-based methods are only suitable for wireless devices exchanging packages frequently but the application scenarios of Shake-n-Shack are for once-off data sharing. Another mainstream of key generation for mobile devices is utilizing the accelerometer readings. For examples, in~\cite{bichler2007key, mayrhofer2009shake}, the authors proposed to hold two mobile devices in one hand and shake them together; therefore the accelerometer reading could be used to generate agreed keys. However, as our previous discussion in Section~\ref{sec:introduction}, these methods are not applicable for smartwatches belonging to different users. The previous work in~\cite{xu2017gait} addressed the problem of automatic key generation for paring the on-body sensor networks by utilizing another natural pattern of human, i.e., gaits. It focused on the mobile devices worn on the same human's body; therefore, it cannot be extended to this work. However,  it is worth noting that we adopt the common quantization and key reconciliation mechanism proposed in previous work, however we solve significantly different problems. 

\noindent \textbf{Data Exchange Between Smart Wearables: }
Finally, there is some work exploring data sharing between smartwatches or waistbands. For examples, the Nabu Smartband is able to share users social contacts via handshake; Apple Inc. drafted a patent~\cite{schorsch2013gesture} about exchanging information between devices in proximity when detecting a ``greeting event'' including handshakes. However, In both of these approaches, handshakes were just regarded as a hint to inform the devices there was a data sharing request, it did not consider to use these handshakes to secure their data sharing process which was the major contribution of this paper. Besides, In a patent~\cite{amento2014devices} submitted by Microsoft Inc., they proposed a new waist-worn hardware which was able to transfer data via human body therefore the secure data sharing could be achieved by physical contacts like handshakes. However, it required extra hardware and could not be adopted by the majority of currently available smart wearables. 
\section{Conclusion}
\label{sec:conclusion}
In this paper, we propose the design and implementation of Shake-n-Shack, a novel system which enables secure data exchange between smart devices via handshakes. The proposed system further blurs the boundaries between the physical and cyber worlds, as it uses the physical contact i.e. handshakes between users, to bridge cyber contact, such as friending on social networks. Under the hood, Shake-n-Shack uses the wrist worn smart wearables, such as smartwatches or fitness bands, to capture the motion patterns induced by handshakes. We show although belonged to different users, that the motion signals of the two hands shaking together are very similar. Based on this, we propose novel approaches to robustly generate and reconcile cryptographic keys on both sides, and use the pair of symmetric keys to establish secure wireless communication channel in a distributed way. We evaluate the proposed Shake-n-Shack system extensively in real-world settings, and experimental results show that the proposed Shake-n-Shack system: a) is able to reliably generate high quality keys within less than two seconds of handshaking; b) is very resilient to the mimicking attacks, achieving only $1.6\%$ Equal Error Rate; and c) can run in real-time on off-the-shelf smartwatches with very slim resource consumption.

\bibliographystyle{IEEEtran}
\bibliography{reference}

\begin{thebibliography}{10}
\providecommand{\url}[1]{#1}
\csname url@samestyle\endcsname
\providecommand{\newblock}{\relax}
\providecommand{\bibinfo}[2]{#2}
\providecommand{\BIBentrySTDinterwordspacing}{\spaceskip=0pt\relax}
\providecommand{\BIBentryALTinterwordstretchfactor}{4}
\providecommand{\BIBentryALTinterwordspacing}{\spaceskip=\fontdimen2\font plus
\BIBentryALTinterwordstretchfactor\fontdimen3\font minus
  \fontdimen4\font\relax}
\providecommand{\BIBforeignlanguage}[2]{{%
\expandafter\ifx\csname l@#1\endcsname\relax
\typeout{** WARNING: IEEEtran.bst: No hyphenation pattern has been}%
\typeout{** loaded for the language `#1'. Using the pattern for}%
\typeout{** the default language instead.}%
\else
\language=\csname l@#1\endcsname
\fi
#2}}
\providecommand{\BIBdecl}{\relax}
\BIBdecl

\bibitem{watch2020}
A.~M. Research, ``Smartwatch market is expected to reach 32.9 billion,
  globally, by 2020,'' \url{https://goo.gl/DJjHYs}, 2016.

\bibitem{apple2017independant}
``Apple newsroom: Watch series 3 brings built-in cellular, powerful new health
  and fitness enhancements.''

\bibitem{rallapalli2014enabling}
S.~Rallapalli, A.~Ganesan, K.~Chintalapudi, V.~N. Padmanabhan, and L.~Qiu,
  ``Enabling physical analytics in retail stores using smart glasses,'' in
  \emph{Proceedings of the 20th annual international conference on Mobile
  computing and networking}.\hskip 1em plus 0.5em minus 0.4em\relax ACM, 2014,
  pp. 115--126.

\bibitem{atzori2010internet}
L.~Atzori, A.~Iera, and G.~Morabito, ``The internet of things: A survey,''
  \emph{Computer networks}, vol.~54, no.~15, pp. 2787--2805, 2010.

\bibitem{mariakakis2016watchudrive}
A.~Mariakakis, V.~Srinivasan, K.~Rachuri, and A.~Mukherji, ``Watchudrive:
  Differentiating drivers and passengers using smartwatches,'' in
  \emph{Pervasive Computing and Communication Workshops (PerCom Workshops),
  2016 IEEE International Conference on}.\hskip 1em plus 0.5em minus
  0.4em\relax IEEE, 2016, pp. 1--4.

\bibitem{sen2016did}
S.~Sen, K.~K. Rachuri, A.~Mukherji, and A.~Misra, ``Did you take a break today?
  detecting playing foosball using your smartwatch,'' in \emph{Pervasive
  Computing and Communication Workshops (PerCom Workshops), 2016 IEEE
  International Conference on}.\hskip 1em plus 0.5em minus 0.4em\relax IEEE,
  2016, pp. 1--6.

\bibitem{wiki:AirDrop}
Wikipedia, ``{AirDrop} --- {W}ikipedia{,} the free encyclopedia,''
  \url{http://en.wikipedia.org/w/index.php?title=AirDrop&oldid=795508835},
  2017, [Online; accessed 20-September-2017].

\bibitem{razer}
\BIBentryALTinterwordspacing
``Razer nabu - social wearable smartband with display and sensor.'' [Online].
  Available: \url{https://www2.razerzone.com/nabu}
\BIBentrySTDinterwordspacing

\bibitem{schorsch2013gesture}
B.~W. Schorsch, D.~J. Shoemaker, E.~Dvortsov, and K.~Kudchadkar,
  ``Gesture-based information exchange between devices in proximity,'' Dec.~18
  2013, uS Patent App. 15/102,834.

\bibitem{diffie1976new}
W.~Diffie and M.~Hellman, ``New directions in cryptography,'' \emph{IEEE
  transactions on Information Theory}, vol.~22, no.~6, pp. 644--654, 1976.

\bibitem{ku2005impersonation}
W.-C. Ku and S.-T. Chang, ``Impersonation attack on a dynamic id-based remote
  user authentication scheme using smart cards,'' \emph{IEICE Transactions on
  Communications}, vol.~88, no.~5, pp. 2165--2167, 2005.

\bibitem{xu2017gait}
W.~Xu, C.~Javali, G.~Revadigar, C.~Luo, N.~Bergmann, and W.~Hu, ``Gait-key: A
  gait-based shared secret key generation protocol for wearable devices,''
  \emph{ACM Transactions on Sensor Networks (TOSN)}, vol.~13, no.~1, p.~6,
  2017.

\bibitem{mayrhofer2009shake}
R.~Mayrhofer and H.~Gellersen, ``Shake well before use: Intuitive and secure
  pairing of mobile devices,'' \emph{IEEE Transactions on Mobile Computing},
  vol.~8, no.~6, pp. 792--806, 2009.

\bibitem{geambasu2009vanish}
R.~Geambasu, T.~Kohno, A.~A. Levy, and H.~M. Levy, ``Vanish: Increasing data
  privacy with self-destructing data.'' in \emph{USENIX Security Symposium},
  vol. 316, 2009.

\bibitem{javali2015secret}
C.~Javali, G.~Revadigar, M.~Ding, and S.~Jha, ``Secret key generation by
  virtual link estimation,'' in \emph{Proceedings of the 10th EAI International
  Conference on Body Area Networks}.\hskip 1em plus 0.5em minus 0.4em\relax
  ICST (Institute for Computer Sciences, Social-Informatics and
  Telecommunications Engineering), 2015, pp. 301--307.

\bibitem{revadigar2015dlink}
G.~Revadigar, C.~Javali, W.~Hu, and S.~Jha, ``Dlink: Dual link based radio
  frequency fingerprinting for wearable devices,'' in \emph{Local Computer
  Networks (LCN), 2015 IEEE 40th Conference on}.\hskip 1em plus 0.5em minus
  0.4em\relax IEEE, 2015, pp. 329--337.

\bibitem{arduser2016recognizing}
L.~Ard{\"u}ser, P.~Bissig, P.~Brandes, and R.~Wattenhofer, ``Recognizing text
  using motion data from a smartwatch,'' in \emph{Pervasive Computing and
  Communication Workshops (PerCom Workshops), 2016 IEEE International
  Conference on}.\hskip 1em plus 0.5em minus 0.4em\relax IEEE, 2016, pp. 1--6.

\bibitem{bhattacharya2016smart}
S.~Bhattacharya and N.~D. Lane, ``From smart to deep: Robust activity
  recognition on smartwatches using deep learning,'' in \emph{Pervasive
  Computing and Communication Workshops (PerCom Workshops), 2016 IEEE
  International Conference on}.\hskip 1em plus 0.5em minus 0.4em\relax IEEE,
  2016, pp. 1--6.

\bibitem{xu2017gaitwatch}
W.~Xu, Y.~Shen, Y.~Zhang, N.~Bergmann, and W.~Hu, ``Gait-watch: A context-aware
  authentication system for smart watch based on gait recognition,'' in
  \emph{Proceedings of the Second International Conference on
  Internet-of-Things Design and Implementation}.\hskip 1em plus 0.5em minus
  0.4em\relax ACM, 2017, pp. 59--70.

\bibitem{revadigar2015mobility}
G.~Revadigar, C.~Javali, H.~J. Asghar, K.~B. Rasmussen, and S.~Jha, ``Mobility
  independent secret key generation for wearable health-care devices,'' in
  \emph{Proceedings of the 10th EAI International Conference on Body Area
  Networks}.\hskip 1em plus 0.5em minus 0.4em\relax ICST (Institute for
  Computer Sciences, Social-Informatics and Telecommunications Engineering),
  2015, pp. 294--300.

\bibitem{revadigar2016secure}
G.~Revadigar, C.~Javali, W.~Xu, W.~Hu, and S.~Jha, ``Secure key generation and
  distribution protocol for wearable devices,'' in \emph{Pervasive Computing
  and Communication Workshops (PerCom Workshops), 2016 IEEE International
  Conference on}.\hskip 1em plus 0.5em minus 0.4em\relax IEEE, 2016, pp. 1--4.

\bibitem{javali2014seak}
C.~Javali, G.~Revadigar, L.~Libman, and S.~Jha, ``Seak: Secure authentication
  and key generation protocol based on dual antennas for wireless body area
  networks,'' in \emph{International Workshop on Radio Frequency
  Identification: Security and Privacy Issues}.\hskip 1em plus 0.5em minus
  0.4em\relax Springer, 2014, pp. 74--89.

\bibitem{shi2013ask}
L.~Shi, J.~Yuan, S.~Yu, and M.~Li, ``Ask-ban: authenticated secret key
  extraction utilizing channel characteristics for body area networks,'' in
  \emph{Proceedings of the sixth ACM conference on Security and privacy in
  wireless and mobile networks}.\hskip 1em plus 0.5em minus 0.4em\relax ACM,
  2013, pp. 155--166.

\bibitem{bichler2007key}
D.~Bichler, G.~Stromberg, M.~Huemer, and M.~L{\"o}w, ``Key generation based on
  acceleration data of shaking processes,'' \emph{UbiComp 2007: Ubiquitous
  Computing}, pp. 304--317, 2007.

\bibitem{amento2014devices}
B.~Amento, K.~A. Li, K.~H. Purdy, and L.~Stead, ``Devices and methods for
  transferring data through a human body,'' Dec.~9 2014, uS Patent 8,908,894.

\end{thebibliography}

\end{document}